\documentclass[aps,prd,twocolumn,nofootinbib]{revtex4-1}
\usepackage{hyperref}
\usepackage[utf8]{inputenc}
\usepackage{textcomp}
\usepackage{graphicx}
\usepackage{t1enc}
\usepackage{amsmath}
\usepackage{mathrsfs}
\usepackage[T1]{fontenc}
\usepackage{amsfonts}
\usepackage{color}
\usepackage{wasysym}
\newcommand{\ud}{\mathrm{d}}
\newcommand{\ui}{\mathrm{i}}

\begin{document}
\title{Laws of black hole thermodynamics in semiclassical gravity}
\author{Bruno Arderucio-Costa}
\email{arderucio@alumni.ubc.ca}
\affiliation{Dept. of Physics and Astronomy, University of British Columbia}

\date{\today}

\begin{abstract}
The first and second laws of black hole thermodynamics are verified to emerge from a generic semiclassical theory of gravity for which a Hamiltonian formulation can be defined. The first law is established for stationary spacetimes, and the second law is established in the here defined ``piecewise stationary'' spacetimes. Black hole entropy is defined in the Noether charge approach, and the entropy for the matter in its exterior is the von Neumann entropy of the quantum fields. These results strongly support the view of black hole entropy as an account of the information that is hidden behind the horizon.
\end{abstract}

\maketitle

\section{Introduction}
Since Hawking's discovery that black holes emit thermal radiation and are consequently assigned with a nonzero temperature, 
the interplay between black hole and thermodynamics gained physical relevance. Because laws of ordinary thermodynamics are presently derived from more basic laws of nature, it is natural to seek a similar derivation for the laws of black hole thermodynamics.

A possible starting point for deriving the the laws of ordinary thermodynamics is Shannon's entropy. It is a functional of a probability density function (PDF), assigning a real positive number that measures the ``amount of ignorance'' about a system. This notion is known to coincide with Boltzmann's in the case the PDF measures the probability density of an ensemble of particles in phase space under certain conditions, and not necessarily appeals to the accessible volume of the phase space that appears in the Boltzmann's original formula. Counting the number of ``degrees of freedom'' does not seem promising in case of black holes, but accounting for the information that is fundamentally inaccessible for observers outside of the black hole seems a natural candidate for it. Since the 1980's \cite{Bombelli} this view is being considered in the literature. If this notion is shown to be correct, our understanding of black hole thermodynamics is elevated to the same status of our understanding of the ordinary thermodynamics.

Here we undertake the reverse task of deriving the laws of black hole thermodynamics from the information perspective and from their derivation interpreting the meaning of black hole's entropy. This work begins by extending the validity of the first law in terms of the Noether charge from a classical theory to a semiclassical version under certain hypothesis. Appendix B contains a review of the Noether charge formalism for classical theories, some familiarity with its methods is advised. Then we prove the generalized second law of thermodynamics (GSL) under a controlled set of assumptions and also using the Noether charge approach. The zeroth law is unchanged moving from a classical theory to a semiclassical theory with a bifurcate Killing horizon (see appendix A for a proof). The results endorse the interpretation of black hole's entropy from the information perspective.

\section{Hamiltonian formalism for semiclassical gravity in stationary spacetimes and the first law}

The set-up is a semiclassical theory $(M,g,\psi,\rho)$ for (bosonic) matter fields $\psi$ that are quantised in a state $\rho$, in the spacetime $(M,g)$, which is treated classically. For the first law, the spacetime is supposed stationary.

We start by separating the degrees of freedom of pure gravity, $g$, and matter, $\psi$, from the general Noether charge formalism reviewed in appendix B. Using the same notation, we write the classical action $I_0=\int\mathbf L(g,\psi,\nabla g,\nabla\psi,\ldots)$ of the theory as a sum 
 $\mathbf L^g+\mathbf L^\psi$, where the first term represents the vacuum contribution, independent of the matter fields, to the Lagrangian form. 
A variation of the Lagrangian can be written as in (\ref{deltaL})
\begin{equation}
 \delta\mathbf L=\mathbf E^\psi\cdot\delta\psi+\mathbf E^g\cdot\delta g+\mathbf R\cdot\delta g+\ud\boldsymbol\Theta^g+\ud\boldsymbol\Theta^\psi,
 \label{Ndell}
\end{equation}
for some form $\mathbf R$. Here $\mathbf E^g=0$ would represent the equations of gravity in absence of matter and $\mathbf E^g+\mathbf R=0$ the full equations of gravity. Requiring the first variation of the Lagragian is zero for any $\delta\psi$ means $\mathbf E^\psi=0$, the equations of motion for the matter. Defining the contributions $\mathbf j_\chi^g$ and $\mathbf j_\chi^\psi$ to the symplectic current as in (\ref{defj}) coming from their respective parts of the Lagrangian and repeating the calculations that led to eqs. (\ref{djay}) and (\ref{j}) for the variation (\ref{Ndell}), we now have
\begin{equation}
 \ud\mathbf j_\chi^\psi=-\mathbf E^\psi\mathsterling_\chi\psi-\mathbf R\mathsterling_\chi g \quad\text{and}\quad \ud\mathbf j_\chi^g=-\mathbf E^g\mathsterling_\chi g,
 \label{Ndjay}
\end{equation}
This means that, remarkably, both contributions $\mathbf j_\chi^\psi$ and $\mathbf j_\chi^g$ to the the symplectic current remain closed if we require the field equations to hold \emph{and} $\mathsterling_\chi g=0$.

We can manipulate the expression for $\ud\mathbf j_\chi^g$ similarly to Iyer and Wald's construction for the symplectic current in a fixed background in \cite{Iyer}:
\begin{multline*}
 \ud\mathbf j_\chi^g=-\boldsymbol\epsilon(E^g)^{ab}(\mathsterling_\chi g)_{ab}=\\
 =-2\boldsymbol\epsilon(E^g)^{ab}\nabla_{(a}\chi_{b)}=-2\boldsymbol\epsilon\{\nabla_a[(E^g)^{ab}\chi_b]+\chi_b\nabla_a(E^g)^{ab}\}=\\
 =-2\ud[(\chi_b(E^g)^{ab})\cdot\boldsymbol\epsilon]+2\chi_b\nabla_a(E^g)^{ab}\boldsymbol\epsilon,
\end{multline*}
where we have written $\mathbf E^g\mathsterling_\chi g$ in terms of the volume form as $(E^g)^{ab}(\mathsterling_\chi g)_{ab}\boldsymbol\epsilon$ and made use of the fact that owing to the symmetry of $g_{ab}$, $(E^g)^{ab}$ can be taken to be symmetric without loss of generality. The left-hand side of the first line is an exact form, and so is the first term of the last line. This means that the second term of the last line has also to be exact. Moreover, this identity has to hold for any field $\chi^a$ at this stage, which is only possible when $\nabla_b(E^g)^{ab}=0$.

This means that, imposing the equations of motion only for the fields, one can integrate the $g$-part of the symplectic current as
\begin{equation}
\mathbf j_\chi^g=-2\boldsymbol\epsilon\cdot(\chi\cdot E^g)+\ud\mathbf Q_\chi^g,
 \label{Njg}
\end{equation}
where $\boldsymbol\epsilon\cdot (\chi\cdot E^g)$ is the shorthand for $(\chi_b(E^g)^{ab})\cdot\boldsymbol\epsilon$, and the integration constant $\mathbf Q_\chi^g$ is the $g$-part of the Noether charge. This can be seen by noticing that if one could impose $\mathbf E^g=0$, eq. (\ref{Njg}) would reduce to the definition of gravitational part of the Noether charge accordingly to the above prescription.

The exact same manipulation for the $\psi$-part of the symplectic current using the first of eq. (\ref{Ndjay}) reveals
\begin{equation}
\mathbf j_\chi^\psi=-2\boldsymbol\epsilon\cdot(\chi\cdot R)+\ud\mathbf Q_\chi^\psi,
 \label{Njp}
\end{equation}
where $R\boldsymbol\epsilon=\mathbf R$ was defined.

Instead of varying the exterior derivative of the current, we can vary the current itself directly from its definition. Following the same steps as in eq. (\ref{j}),
\begin{multline}
\boldsymbol\omega_\chi^\psi=\delta\mathbf j_\chi^\psi-\ud(\chi\cdot\boldsymbol\Theta^\psi)+\chi\cdot\mathbf R\delta g\\
\text{and}\\
\boldsymbol\omega_\chi^g=\delta\mathbf j_\chi^g-\ud(\chi\cdot\boldsymbol\Theta^g)+\chi\cdot\mathbf E^g\delta g,
 \label{newj}
\end{multline}
with the forms $\boldsymbol{\omega}_\chi^{\psi,g}$ constructed from the symplectic potentials $\boldsymbol{\Theta}^{\psi,g}$ as in (\ref{sfgen1}) and (\ref{sfgen2}). For the last equation, we suppose both $\psi$ and $\psi+\delta\psi$ obey $\mathbf E^\psi=0$.

As expected when the equations of motion for gravity $\mathbf E^g+\mathbf R=0$ are satisfied, $\boldsymbol\omega_\chi=\boldsymbol\omega_\chi^\psi+\boldsymbol\omega_\chi^g$ and $\ud\boldsymbol\omega_\chi=0$.

Can one exploit this formulation in a semiclassical context? To address this question, I start with a toy model of as a warm-up. Consider the Lagrangian
\[\mathbf L(q,Q,\dot q,\dot Q)=\left\{\frac{1}{2}(\dot q^2+\dot Q^2)-\frac{1}{2}(Q^2+2c\ qQ)\right\}\ud t,\]
of which only the first term represents the analogue of $L^g$, the other three terms compose $L^\psi$. In this expression $c$ is a coupling constant. From a variation of the above
\begin{multline*}
    \delta\mathbf L=-\left[\underbrace{(\ddot Q+Q+cq)}_{E^\psi}\delta Q+\underbrace{\ddot q}_{E^g}\delta q+\right.\\
    \left.\underbrace{cQ}_R\delta q\right]\ud t+\ud\left(\underbrace{\dot q\delta q}_{\Theta^g}+\underbrace{\dot Q\delta Q}_{\Theta^\psi}\right)
\end{multline*}
we obtain the symplectic current of the field $\xi=\partial/\partial t$
\[j_\xi^\psi=\frac{1}{2}[\dot Q^2+\dot q^2+Q^2+c\ qQ].\]
And using the first of eqs. (\ref{newj}) and the analogue of eq. (\ref{dh}),
\[\delta H_\xi^\psi
=Q\delta Q+P\delta P+cq\delta Q,\]
where $p=\dot q$ and $P=\dot Q$.

We can now promote $Q$ and $P$ to operators. In our example $(P,Q)$ is a symplectic coordinate system. Hence, if the (classical) Hamiltonian function is evaluated at the corresponding operators $P$ an $Q$ to obtain the quantum Hamiltonian operator, the same Hamilton's equations, now interpreted as operator-valued, are satisfied\footnote{Explicitly, the quantum equations of motion in the Heisenberg picture for the Hamiltonian operator $H^\psi=\frac{1}{2}\left(Q^2+P^2\right)+cqQ$ are
\begin{multline*}
    \dot Q=-\ui[Q,H^\psi]=P=\delta H^\psi/\delta P\\
    \text{and}\\
    \dot P=-\ui[P,H^\psi]=-Q-cq\mathbf1=-\delta H^\psi/\delta Q,
\end{multline*} as claimed.}.

Satisfying the operator-valued Hamilton's equations is not a particularity of our toy-model Lagrangian. In general, the formal relations $\ui[f(p_i),q_k]=-\partial f/\partial p_i\ \delta_{ik}$ and $\ui[f(q_i),p_k]=\partial f/\partial q_i\ \delta_{ik}$\footnote{These relations can be verified when $f$ is a power of its argument by induction, and subsequently generalised for other functions by making use of the density property of polynomials over the set of continuous functions.} obeyed by the operators $\{(p_i,q_i)\}_i$ associated with a symplectic coordinate system ensure that Heisenberg's equations of motion are equivalent to the operator-valued Hamilton's equations at any given time.

It is justifiable to write a semiclassical 
Hamiltonian of the $g$-part of this interacting theory formally as
\begin{equation}
 \delta H_\xi^g=p\delta p+c\langle Q\rangle\delta q
 \label{toyDh}
\end{equation}
since it generates the desired Hamilton's equations for the classical degrees of freedom $\delta H_\xi/\delta p=p$ and $\delta H_\xi/\delta q=c\langle Q\rangle$. And overall, we can interpret $\boldsymbol\omega_\xi=\boldsymbol\omega_\xi^g+\langle\boldsymbol\omega_\xi^\psi\rangle$ as the pre-symplectic form generating the full semiclassical theory. Variations $\delta$ of functions of operators are understood as ``derivatives'' like in their classical counterparts.

Returning to a more general theory, I consider, for simplicity, only quadratic, free fields, to avoid dealing with counterterms in the action and with altered equations of motion. The Hamiltonian operator $\delta H_\chi^\psi$ whose flow is generated by $\chi^a$ for the quantum fields is obtained from the classical function by preserving the format of $\mathbf j_\chi^\psi$, $\boldsymbol\Theta^\psi$, and $\mathbf R$ as functions of the fields to be evaluated on the field operators and then take their regularised expectation values $\langle\mathbf j_\chi^\psi\rangle$, $\langle\boldsymbol\Theta^\psi\rangle$, and $\langle\mathbf R\rangle$ respectively and integrating the first of eq. (\ref{newj}) over a Cauchy surface $C$. Because $\boldsymbol\omega_\chi^\psi$ is not closed, the existence of the Hamiltonian is a stricter condition than the existence of the full Hamiltonian of the classical theory. Like in the toy-model, the classical degrees of freedom respond according to 
\begin{equation}
\boldsymbol\omega_\chi^g=\delta\mathbf j_\chi^g-\ud(\chi\cdot\boldsymbol\Theta^g)-\chi\cdot\langle\mathbf R\rangle\delta g,
\label{classresp}
\end{equation}
which is requiring that the semiclassical equations of motion $\mathbf E^g+\langle\mathbf R\rangle=0$ are satisfied in eq. (\ref{newj}).

This construction requires that the semiclassical couples gravity to expectation values of the fields as if they were classical objects, and that the matter field operators satisfy their Hamilton equations under their expectation values for a state $\rho$. When there exists such $(M,g,\psi,\rho)$ for this semiclassical theory, imposing $\mathbf E^\psi=\mathbf E^g+\langle\mathbf R\rangle=0$ produces, unlike usual quantum field theory in a fixed background, a self-consistent theory\footnote{In other words, back reaction of phenomena originated in the quantum mechanical nature of matter are automatically accounted for.}. So kinematic properties of the Killing horizon like the constancy of the surface gravity throughout $\mathfrak h_\pm$ (see appendix A) are automatically true. 


An example of the separation of the matter and gravity degrees of freedom for a concrete theory is given in the appendix B. 

The first law of black hole thermodynamics is obtained as follows. For asymptotically flat stationary spacetimes and stationary metric perturbations, $\boldsymbol\omega_\chi^g=\mathsterling_\chi\boldsymbol\Theta^g(g,\delta g)-\delta\boldsymbol\Theta^g(g,\mathsterling_\chi g)=0$. Then, from eq. (\ref{classresp}),
\begin{equation}
0=\ud\delta\mathbf Q_\chi^g+2\delta[\boldsymbol\epsilon\cdot(\chi\cdot\langle R\rangle)]-\ud(\chi\cdot\boldsymbol\Theta^g)-\chi\cdot\langle\mathbf R\rangle\delta g,
 \label{pre1}
\end{equation}
where we have used eq. (\ref{Njg}) and $E^g=-\langle R\rangle$.

We now integrate eq. (\ref{pre1}) over a spacelike surface $C$ bounded by the bifurcation surface $B$ and a $(n-2)$-sphere at spatial infinity, denoted by $\infty$. We can eliminate the terms containing $\langle\mathbf R\rangle$ in favour of the regularised expectation value of the Hamiltonian $\langle\delta H_\chi^\psi\rangle=\int_C\langle\boldsymbol\omega_\chi^\psi\rangle$ of the field theory defined over the domain of dependence of $C$. Explicitly, from eqs. (\ref{newj}) and (\ref{Njp}),
\begin{multline*}
    \int_C \left\{2\delta[\boldsymbol\epsilon\cdot(\chi\cdot\langle R\rangle)] -\chi\cdot\langle\mathbf R\rangle\delta g\right\}=\\
    =-\langle\delta H^\psi_\chi\rangle+\int_C\left\{\ud\delta\langle Q_\chi^\psi\rangle-\ud(\chi\cdot\langle\boldsymbol\Theta^\psi\rangle)\right\}.
\end{multline*}

Plugging into eq. (\ref{pre1}),

\begin{equation}
 -\langle\delta H_\chi^\psi\rangle+\int_{\partial C}\left\{\delta\mathbf Q_\chi-(\chi\cdot\boldsymbol\Theta)\right\}=0,
 \label{main1}
\end{equation}
where $\mathbf Q_\chi=\mathbf Q_\chi^g+\langle\mathbf Q_\chi^\psi\rangle$, $\boldsymbol\Theta=\boldsymbol\Theta^g+\langle\boldsymbol\Theta^\psi\rangle$, and Stokes' theorem has been used.

The right-hand side of eq. (\ref{main1}) is evaluated as an integral over $B$ plus an integral over $\infty$. Over $B$, the second term vanished and, for a black hole with constant surface gravity, the first is simply the variation of Noether charge entropy $S_\text{NC}$ multiplied by $-\kappa/2\pi$ \cite{Wald,Iyer}. And the integral over $\infty$ is (assuming the existence of a form $\mathbf B$ such that $\boldsymbol\Theta|_\infty=\delta\mathbf B$) the variation of the canonical quantity $\mathcal E_\chi$ conjugated to the Killing parameter of $\chi$, i.e., the corresponding ADM conserved quantity \cite{Wald99} (``conserved'' in the sense of not depending on the choice of the integration three-surface). Thus,

\begin{equation}
\frac{\kappa}{2\pi}\delta S_\text{NC}+\langle\delta H_\chi^\psi\rangle=\delta\mathcal E_\chi.
\label{main2}
\end{equation}

Equation (\ref{main2}) is already a form of the first law of black hole thermodynamics. But we can also write it in a different way.

In some applications it is useful to understand expectation value of variations of operators, say $\delta H_\chi^\psi$, in the Sch\"odinger picture instead of Heisenberg's. If 
there is a unitary map $\mathscr U$ between the Heisenberg operators so that $\mathcal O\rightarrow\mathcal O+\delta\mathcal O=\mathscr U\mathcal O\mathscr U^\dagger$, in Schr\"odinger picture, we then keep the operators unchanged and produce the same expectation values $\langle\delta\mathcal O\rangle$ by transforming the state $\rho$ by $\rho\rightarrow\mathscr U^\dagger\rho\mathscr U$.

Adopting this picture, we investigate the case where the unperturbed state defined over the right wedge (i.e., the domain of dependence of $C$) is the Hartle-Hawking state given by

\begin{equation}
 \rho_0=\frac{1}{Z}\exp\left(-\frac{2\pi}{\kappa}H_\chi^\psi\right),
 \label{density}
\end{equation}
where $H_\chi^\psi$ is the Hamiltonian operator with respect to the ``time translation'' defined by $\chi^a$ as above, and $1/Z$ is a normalisation factor. Strictly speaking, this density matrix is only defined when the spectrum of $\hat H_\chi$ is discrete, so that $Z=\mathrm{Tr}\ e^{-2\pi H_\chi/\kappa}$ is a finite quantity, and this problem is normally bypassed by confining the field in a box with certain boundary conditions. This is not a problem of fundamental physical significance, and thermal states can be defined precisely through the Kubo-Martin-Schwinger condition \cite{KayWald}.

The difference between von Neumann entropies associated with the states $\rho_0$ and $\rho_0+\delta\rho$ is given by
\begin{multline}
     \delta S_\text{vN}=\mathrm{Tr}(\rho_0\log\rho_0)-\mathrm{Tr}[(\rho+\delta\rho)\log(\rho_0+\delta\rho)]=\\
     -\mathrm{Tr}(\rho_0\rho_0^{-1}\delta\rho)-\mathrm{Tr}(\delta \rho\ \log\rho_0)+O(\delta\rho^2)\\
     =\frac{2\pi}{\kappa}\langle\delta H_\chi^\psi\rangle+O(\delta\rho^2),
 \label{main}
\end{multline}
where we used that fact that normalisation of both $\rho_0$ and $\rho_0+\delta\rho$ implies $\mathrm{Tr}\ \delta\rho=0$ and used (\ref{density}) in the last step.

Combining eqs. (\ref{main}) and (\ref{main2}) we get a statement of the first law in terms of the entropy of the matter:
\begin{equation}
 \frac{\kappa}{2\pi}\delta(S_\text{vN}+S_\text{NC})=\delta\mathcal E_\chi.
 \label{1stlaw}
\end{equation}

\section{Generalized second law}
Differently from the last section, results concerning the generalised second law of thermodynamics (GSL) have to be established in non-stationary spacetimes,
 where the notion of black hole entropy as Noether charge is not so solidly established. For stationary black holes, classical fields and stationary perturbations, it has been shown by Jacobson, Kang, and Myers \cite{Ted} that the integral of the Noether charge does not depend on the integration surface, in particular does not need to be evaluated at the bifurcation surface. It is a local geometrical object and all the dependence on the Killing vector tangent to the horizon can be eliminated on the bifurcation surface as follows: its value restricted on the surface is $\chi^a=0$, its first derivative is written in terms of the volume form (\ref{volformB}), and any higher derivatives can be eliminated by applying successively the Killing identity $\nabla_c\nabla_a\chi_b=-R_{abcd}\chi^d$. For non-stationary black holes on the other hand, or even for stationary black holes but non-stationary perturbations, these properties are lost.
 
 Iyer and Wald \cite{Iyer} listed the following desirable conditions the entropy of dynamical black holes ought to satisfy: (\emph{i}) coincides with the Noether charge entropy in any cross section for stationary black holes, (\emph{ii}) its first variation evaluated on the bifurcation surface obeys a version of the first law of black hole mechanics, (\emph{iii}) is invariant by altering the Lagrangian by adding an exact form, and (\emph{iv}) obeys a version of the second law of the black hole mechanics. 
 
 They proposed a definition satisfying the first three conditions, but not the fourth. The proposal here obeys (i), (ii) and (iv) corrected for effects of semiclassical origin, but fails to obey (iii). Fortunately, relaxing condition (iii) is not problematic since, although theories with identical dynamical contents can generate different entropies, arguments coming from the path integral approach to quantum gravity suggest that the boundary terms on the Lagrangian can be physically meaningful for a future theory. For example, Hawking\cite{nosp, grecs} considers that the transition amplitude from a state $|\psi,h_{ab}\rangle$ to a state $|\psi',h'_{ab}\rangle$ should be given by a Feynman integral over fields $\psi$ and induced metrics $h_{ab}$. If this amplitude is to factor out $\langle\psi, h_{ab}|\psi^\prime,h_{ab}^\prime\rangle=\sum_{\psi^{\prime\prime}h^{\prime\prime}}\langle\psi, h_{ab}|\psi^{\prime\prime}h_{ab}^{\prime\prime}\rangle\langle\psi^{\prime\prime}h_{ab}^{\prime\prime}|\psi^\prime,h_{ab}^\prime\rangle$ when an intermediate state $|\psi^{\prime\prime}h_{ab}^{\prime\prime}\rangle$ is placed, the classical actions must add. For the Einstein-Hilbert action, this only happens when a boundary term proportional to the trace of the extrinsic curvature is included in the action (see, for example, appendix E of \cite{WaldGR}). This term is, apart from an additive constant, referred to as the Hawking-Gibbons-York term in the action.
 
 We consider the semiclassical theory $(M,g,\psi,\rho)$ consisting of a future-predictable, asymptotically flat, ``piecewise stationary'' (definition below) spacetime $(M,g)$ and consistent matter fields obeying the semiclassical equations $\mathbf E=0$\footnote{In our notation from previous section, this denotes satisfying both $\mathbf E^\psi=0$ and $(\mathbf E^g+\langle\mathbf R\rangle)=0$}. Intuitively speaking, ``piecewise stationarity'' means that the spacetime behaves like a stationary one up until a surface $S_1$, evolves dynamically up until a second surface $S_2$, when it becomes stationary again.
 
 More rigorously, a spacetime is called \emph{piecewise stationary} if there are two Cauchy surfaces $S_1$ and $S_2\subset I^+(S_1)$ such that $J^-(S_1)$ is isometric (with isometry $\varphi_1$) to $J^-(S_1')\subset M_1$ and $J^+(S_2)$ to $J^+(S_2')\subset M_2$ ($\varphi_2$ denotes the isometry), where $(M_1, g_1)$ and $(M_2, g_2)$ are (unphysical) stationary spacetimes and $S'_1$ and $S'_2$ are Cauchy surfaces in their respective manifolds.
 
The case of interest is when there is a black hole on $(M,g)$. Being a system with negative heat capacity, black hole can only be in a ``stable thermal equilibrium'' with an extensive system when the latter has its energy bounded, so that if the extensive system gains (loses) energy to the black hole, its temperature increases (decreases) faster than the black hole's \cite{Hawking76}. An alternative way of interpreting the impossibility of a stable thermal equilibrium in a system with a black hole in an infinite reservoir is picturing a localised perturbation which requires an infinite amount of time to get to infinity, so an originally stationary spacetime cannot become stationary again in a finite amount of time.

\begin{figure}[h]
      \centering
      \includegraphics[width=.85\columnwidth]{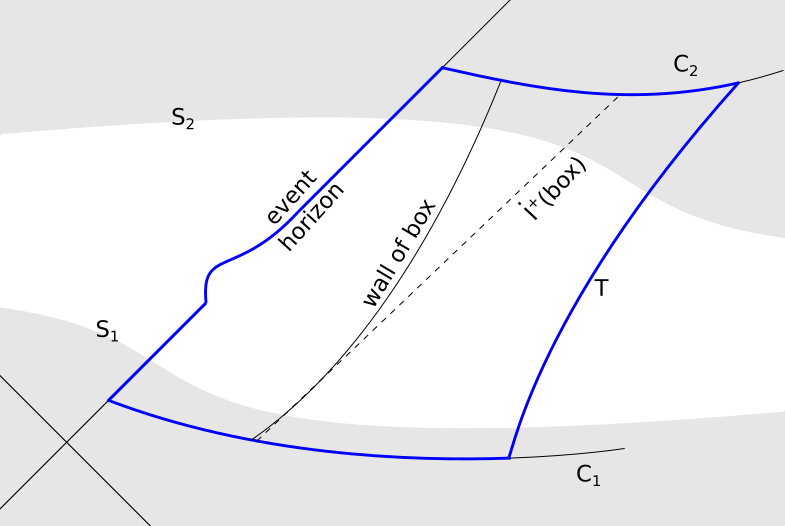}
      \caption{Geometry that leads to eq. \ref{bigint}. The worldline labeled as ``wall of box'' denotes the timeline along which the boundary conditions on the fields are imposed. The shaded areas represent the regions of $M$ that are isometric to stationary spacetimes. The contour of integration is in blue.}\label{GSLconstruction}
  \end{figure}

Thus, in order to make the hypothesis of piecewise stationarity attainable, we confine the fields $\psi$ in a ``box'' on whose walls we impose as a boundary condition that any component of $\langle\star\mathbf j_\chi^\psi\rangle$ orthogonal to them must vanish (i.e., no energy can flow into or out of the box). The walls are outside the black hole in the region between $S_1$ and $S_2$ (as in figure \ref{GSLconstruction}). We further assume that all the disturbance responsible for the breakdown of stationarity comes from localised changes in the states of the fields within the box.

Let $\chi_1^a$ denote a generator of the event horizon on $(M_1,g_1)$. This field can seen as a field in the physical spacetime $J^-(S_1)$ through the isometry, $\chi^a=\varphi_{1*}^{-1}\chi_1^a$, as a generator of the event horizon in $(M,g)$. As any generator of the event horizon, it is future-inextendible and entirely contained in the event horizon of $M$ \cite{HE}. In particular, it generates the event horizon in $J^+(S_2)$, where it can be used to define the field $\chi_2^a=\varphi_{2*}\chi^a$ on $(M_2,g_2)$ which generates the event horizon of $(M_2,g_2)$. According to the rigidity theorem (see Proposition 9.3.6 of ref. \cite{HE}), $\chi^a_2$ generates a \emph{Killing} horizon in $(M_2,g_2)$.
 
 In contrast, although there is no guarantee that $\chi_1^a$ will be a Killing field, this is the case in the limit where the horizon approaches $i^-$. In this limit, the event horizon and the apparent horizon coincide. The former has zero shear and vorticity, and the latter zero expansion, so in that limit $\chi_1^a$ obeys the Killing equation and is null. In other words, the event horizon approaches a Killing horizon generated by $\chi_1^a$. This fact can be illustrated using an ingoing Vaidya metric\footnote{This spacetime is not strictly speaking piecewise stationary, but it serves to the purpose of illustrating the coincidence of the event and apparent horizon.} $\ud s^2=-(1-2m(v)/r)\ud v^2+2\ud v\ud r+r^2\ud\Omega^2$, with $m(v)=m_1>0$ in the past of a surface of constant $v=v_0$, and $m(v)=m_2>m_1$ on its future. In the past of the surface $v=v_0$, the event horizon is the surface of constant $u=v-2r^*, \ r^*=r+2m_1\log(r-2m_1)$ which coincides with $r=2m_2$ on $v=v_0$. This surface approaches the apparent horizon $r=2m_1$ exponentially fast as one moves to the past. For a solar-mass object, the characteristic Schwarzschild time $2m_1$ is of order $10^{-5}$ seconds.

 The initial data of the field modes $\phi$ in $S'_1$ is identified with the pullback of modes $\varphi^*_1\phi$ on $S_1$. In other words, we choose the modes $\phi$ that solve the field equations on $M$ and whose restriction on $J^-(S_1)$ coincides with the $\varphi^*_1\phi$ and we carry the quantisation process using these modes. Similarly, we can use the isometry $\varphi_2$ to identify modes on $J^+(S_2)$ with modes on $J^+(S'_2)$ and quantise the fields on $M_2$.
 
 If the semiclassical equations of motion hold on the entirety of the spacetime, $\boldsymbol\omega_\chi=\delta\mathbf j_\chi^g+\langle\delta\mathbf j_\chi^\psi\rangle+\ud[\chi\cdot(\boldsymbol{\Theta}^g+\langle\boldsymbol{\Theta}^\psi\rangle)]$ is closed:
 \begin{equation}
  \ud\boldsymbol\omega_\chi=\mathsterling_\chi\mathbf E\delta\phi-\delta\mathbf E\mathsterling_\chi\phi=0,
  \label{omegaclosed}
 \end{equation}
 meaning that the integral of $\boldsymbol\omega_\chi$ over the a closed three-surface is zero.
 
 We choose this surface to be $C_1\cup C_2\cup T\cup h$, where $C_{1,2}$ are portions of two Cauchy surfaces $\Sigma_{1,2}$ of $(M,g)$ outside the black hole, with $\Sigma_2\subset J^+(S_2)$, $\Sigma_1\subset J^-(S_1)$. Later $\Sigma_1$ will be pushed down to the past so that it intercepts the horizon near $i^-$. The segment $h$ is the intervening portion of the event horizon between $\Sigma_1$ and $\Sigma_2$, and $T$ is a timelike surface outside the causal future of the box in which the fields are confined, so that any point of $T$ is spacelike separated to any point within the box, wherein the disturbance in stationarity happens. With this choice, $T$ does not contribute to the integral of exact forms for causal theories\footnote{Using Stoke's theorem, integration of any exact form over $T$ can be converted into integration over $T\cap C_1$ and $T\cap C_2$, where the integrands must share a common value.}. See figure (\ref{GSLconstruction}) for a scheme of this construction.

 The integral becomes
\[
  \left(-\int_{C_1}+\int_{C_2}-\int_h\right)\boldsymbol\omega_\chi=0.
\]
 
 The integrals over $C_1$ and $C_2$ can be evaluated from the isometries, and $\boldsymbol\omega_\chi^g=0$ due to stationarity, like in the previous section, leaving
 \begin{equation}
  \int_h\boldsymbol\omega_\chi=\Delta\langle\delta H_\chi^\psi\rangle,
  \label{bigint}
 \end{equation}
where $\Delta$ denotes the difference in a quantity caused by the change in the integration surface from a domain in $\Sigma_1$ to its corresponding domain in $\Sigma_2$, i.e., a change ``in time''.

A direct calculation of a variation in the full symplectic current, or the addition of $\boldsymbol\omega_\chi^\psi$ and $\boldsymbol\omega_\chi^g$ in eq. (\ref{newj}) together with $\mathbf E^g+\langle\mathbf R\rangle=0$ reveals
\[\boldsymbol\omega_\chi=\delta\mathbf j_\chi-\ud(\chi\cdot\boldsymbol\Theta).\]
Furthermore, if both $(g,\psi)$ and $(g+\delta g,\psi+\delta\psi)$ satisfy the equations of motion $\mathbf E=0$, then adding eqs. (\ref{Njg}) and (\ref{Njp}),
\[\delta\mathbf j_\chi=\delta\ud\mathbf Q_\chi\]
and, under these hypotheses, both $\delta\mathbf j_\chi$ and $\boldsymbol\omega_\chi$ are closed and the integral over $h$ in eq. (\ref{bigint}) can be converted into a boundary integral,
\[\int_h\boldsymbol\omega_\chi=\left(\int_{C_2\cap h}-\int_{C_1\cap h}\right)\left\{\delta \mathbf Q_\chi-\chi\cdot\boldsymbol\Theta\right\},\]

 If both $(M_{1,2},g_{1,2})$ posses a bifurcation surface, the second term does not contribute to the integral. To see this, write $\displaystyle\left(\int_{\varphi_{1,2}(C_{1,2}\cap h)}-\int_{B_{1,2}}\right)\chi\cdot\boldsymbol\Theta=\int_{\mathfrak h'}\chi\cdot\boldsymbol\Theta$, where $\mathfrak h'$ is the intervening portion of the Killing horizon between $\varphi_{1,2}(C_{1,2}\cap h)$ and the bifurcation surfaces $B_{1,2}$. But $\chi$ is tangent to $\mathfrak h'$, so the right-hand side is zero, showing that the original integrals can have their domains of integration replaced by $B_{1,2}$, where $\chi^a=0$.
 
 Putting these elements together, eq. (\ref{bigint}) can be written as
 
 \begin{equation}
 \Delta\langle\delta H_\chi^\psi\rangle=-\Delta\int\delta\mathbf Q[\chi].
 \label{bigdeltas}
\end{equation}

It is well known \cite{Wald, Iyer, Ted} that the decomposition in the Noether charge is not unique, but rather carries ambiguities coming from three different sources, namely, the addition of an exact form in the Lagrangian, which although does not alter the classical dynamical content of the theory, is believed to be important in the quantum theory (as argued above); the addition of an exact form in the Noether charge itself, which since it does not contribute for its integral, is immaterial both classically and quantum mechanically; and finally the addition of an exact form in the current $\mathbf j'=\mathbf j+\ud\mathbf Y$, whose presence is entirely expected, since no requirement to the absolute value of the Hamiltonian was imposed, only to its variations. We can physically interpret this last ambiguity as the addition of a ``second conserved current'' that obeys a separate continuity equation, and therefore is completely independent of the conservation of $\mathbf Q$. This ambiguity is also present on the definition of the expectation value of the Hamiltonian of the quantum fields and comes as no surprise since no particular renormalisation procedure has been specified. If a renormalisation process uniquely defines $\langle H_\chi^\psi\rangle$, a prescription for determining $\mathbf Y$ will follow from it\footnote{For example, Wald's axioms \cite{WaldB} define $\langle T_{ab}\rangle_\text{reg}$ only up to a local geometric function. This means that the classical ambiguity of the presence of the form $\mathbf Y$ manifests itself in the same way as the quantum ambiguity one has for the renormalisation process.}. For this reason, after fixing $\mathbf Y$ and the boundary term in the action, we can define $$S_\text{NC}\equiv\frac{2\pi}{\kappa}\int\mathbf Q[\chi]$$ uniquely in the semiclassical theory. Nevertheless, as it will be clear below, a ``generalised second law'' holds for each choice of $\mathbf Y$. In terms of the entropy, eq. (\ref{bigdeltas}) reads
\begin{equation}
 \Delta\langle\delta H_\chi^\psi\rangle=-\Delta\left[\frac{\kappa}{2\pi}\delta S_\text{NC}.\right]
 \label{bigdeltasp}
\end{equation}

In the density matrix approach, each expectation value of $H_\chi$ can be written as $\mathrm{Tr}(\rho_{1,2}H_\chi)$, where $\rho_{1,2}$ is obtained by taking the partial trace over degrees of freedom lying in $S_{1,2}-C_{1,2}$. The partial trace is a linear positive map between operators defined on their respective spaces, so the theorem by Lindblad in ref. \cite{Lindblad} can be applied, which states that the relative entropy between two states $\rho$ and $\sigma$ $S(\rho|\sigma)\equiv\mathrm{Tr}(\rho\log\rho)-\mathrm{Tr}(\rho\log\sigma)$ cannot increase after the operation is used to obtain $\rho_\text{red}$ and $\sigma_\text{red}$, symbolically
\begin{equation}
S(\rho_\text{red}|\sigma_\text{red})\leq S(\rho|\sigma),
\label{ineq}
\end{equation}
where the suffix ``red'' denotes the reduced density matrix obtained after the partial trace.

Consistently with the hypothesis that $(M,g)$ is stationary in $J^-(S_1)$, we prepare the fields in a state $\sigma_0$, annihilated by the annihilation operators  corresponding to the eigenmodes of $\mathsterling_\xi$, with $\varphi_{1*}\xi^a$ is an asymptotically timelike Killing field of $(M_1,g_1)$. Let $\sigma_1$ denote the reduced density matrix obtained from $\sigma_0$ by tracing out degrees of freedom lying in $S_1-C_1$. All the observables locally constructed on $J^-(S_1)$ can be calculated via $J^-(S'_1)$, where the eigenmodes above diagonalise the Hamiltonian. Hence $\sigma_1$ is described as a Gibbs state associated with $H_\chi^\psi$ and temperature $\kappa_1/2\pi$, $\exp(-2\pi H_\chi^\psi/\kappa_1)/Z_1$, i.e., the corresponding Hartle-Hawking state.

The evolution of $\sigma_0$ in $M$ is supposed to be Hamiltonian, then modes evolve along the orbits of $\xi^a$ according to $\sigma_0\to U^\dagger\sigma_0U$, with $U=\exp(\ui H_\xi^\psi t)$, with $\xi=\partial/\partial t$. To be consistent with the piecewise stationarity of the unperturbed solution, we conclude that in $I^+(S_2)$, the Hamiltonian has to be a constant, so $U$ can be written as $U=\exp(\ui\tilde\omega t)$ for some constant $\tilde\omega$\footnote{This can be seen from the non-degeneracy property of the symplectic form. The evolution of a scalar quantity $F$ is governed by $\ud F/\ud t=\boldsymbol\Omega(\nabla F,\nabla H)$, which is zero for a generic $F$ when $\delta H=0$.}. The density matrix $\sigma_2$ is obtained after tracing out the degrees of freedom lying on $S_2-C_2$ of $\sigma_0$. Again, $\sigma_2$ viewed from $M_2$ is a Gibbs state associated with the Hamiltonian $H_\chi$ and temperature $\kappa_2/2\pi$.

From the definition of a black hole, all causal curves intercepting $S_1-C_1$ necessarily intercept $S_2-C_2$, but since the black hole is dynamical, there are curves intercepting $S_2-C_2$ that do not intercept $S_1-C_1$, meaning that $\rho_2$ can be obtained by reducing $\rho_1$, not the reverse. This is how an arrow of time is introduced in the GSL, it comes from the time-asymmetric nature of black holes.

If $\rho_{1,2}$ is the physical (mixed) state on the algebras defined over $C_{1,2}$, a simple evaluation shows
\begin{equation}
S(\rho_{1,2}|\sigma_{1,2})=\frac{2\pi}{\kappa_{1,2}}(\langle H_\chi\rangle_{\rho_{1,2}}+\log Z_{1,2})-S_\text{vN},
\label{relent}
\end{equation}
and as we know from standard techniques
\begin{equation}
-\frac{\kappa_{1,2}}{2\pi}\log Z_{1,2}=F[\sigma_{1,2}]=\langle H\rangle_{\sigma_{1,2}}-\frac{\kappa_{1,2}}{2\pi}S_\text{vN}[\sigma_{1,2}].
 \label{helmholtz}
\end{equation}

Combining eqs. (\ref{relent}) and (\ref{helmholtz}),
\begin{multline*}
    S(\rho_{1,2}|\sigma_{1,2})=\frac{2\pi}{\kappa_{1,2}}\left(\langle H_\chi\rangle_{\rho_{1,2}}-\langle H_\chi\rangle_{\sigma_{1,2}}\right)\\
    -\left(S_\text{vN}[\rho_{1,2}]-S_\text{vN}[\sigma_{1,2}]\right),
\end{multline*}
together with (\ref{ineq}) and (\ref{bigdeltasp}),
\begin{equation}
\Delta\delta(S_\text{vN}+S_\text{NC})\geq0.
\label{vargsl}
\end{equation}

I emphasise the definition $S_\text{NC 1,2}=\frac{2\pi}{\kappa_{1,2}}\int\mathbf Q[\chi]$ is only evaluated in the regions where the spacetime is stationary, i.e., where the all the previous discussions apply. Any definition of black hole's entropy which agrees with the Noether charge's in stationary spacetimes ought to obey eq. (\ref{vargsl}).

If one can show that the GSL is true for the state $\sigma_0$ then it follows from eq. (\ref{vargsl}) that it is true for any state described as a linear perturbation around it. Two ingredients are needed to complete this task.

First, from the closure of full symplectic current when the full semiclassical equations of motion are satisfied everywhere, an integral of $\mathbf j_\chi$ over the same circuit of figure (\ref{GSLconstruction}) reveals
\begin{equation}
\Delta\langle H_\chi^\psi\rangle_{\sigma_0}+\Delta\int_{h\cap C_{1,2}}\mathbf Q_\chi+\Delta\int_{C_{1,2}}\mathbf j_\chi^g=0,
 \label{1stIngTS}
\end{equation}
where we put $\langle H_\chi^\psi\rangle=\int_{C_{1,2}}[\langle\mathbf j_\chi^\psi\rangle-\ud(\chi\cdot\langle\mathbf B^\psi\rangle)]$ and made use of the fact that $\int_{h\cap C_{1,2}}\chi\cdot\langle\mathbf B^\psi\rangle=0=\Delta\int_{C_{1,2}\cap T}\chi\cdot\langle\mathbf B^\psi\rangle$, the first equality obtained from arguments similar to the ones above eq. (\ref{bigdeltas}).

Second, for the first time, arguments coming from thermodynamics are invoked. More specifically, we suppose that variations of the von Neumann entropy of the open system behave as variations of its thermodynamic entropy. For this system, $\langle H_\chi^\psi\rangle$ cannot be interpreted as a conserved energy. For even if the integration of $\langle \mathbf j_\chi^\psi\rangle-\ud(\chi\cdot\langle\mathbf B^\psi\rangle)$ is taken over entire Cauchy surfaces, its value will still depend on the choice of the integration surface because $\ud\langle\mathbf j_\chi^\psi\rangle\neq0$, meaning that the matter degrees of freedom over a Cauchy surface constitute an open system.

Hence, the first law of ordinary thermodynamics applied to the exterior of the black hole is
\begin{equation}
T\ud S_\text{vN}=\ud\langle H_\chi^\psi\rangle+\ud\int_C \mathbf j_\chi^g,
\label{2ndIngTS}
\end{equation}
 the last term is added to represent the amount of ``conserved energy'' leaves the matter degrees of freedom over $C_{1,2}$ to the gravitational ones over the same surface, $\ud (\langle\mathbf j_\chi^\psi\rangle+\mathbf j_\chi^g)=0$ used. And since the variation $\Delta S_\text{vN}$ of the entropy of the matter field is a function of state, it is the same as in any process that share the same initial and final states for the matter, including one passing through infinitely many equilibrium states with temperature $T$ varying between $\kappa_1/2\pi$ and $\kappa_2/2\pi$ in an ensemble where macrostates are distinguishable by the the expectation value of the Hamiltonian $\langle H_\chi\rangle$. A direct integration of eq. (\ref{2ndIngTS}) gives
\begin{equation}
\Delta S_\text{vN}=-\Delta S_\text{NC},
\label{saturation}
\end{equation}
 after using eq. (\ref{1stIngTS}) and the definition of the Noether charge entropy.
 
 Putting this equality and eq. (\ref{vargsl}) together we have the GSL
\begin{equation}
\Delta (S_\text{vN}+S_\text{NC})\geq0,
 \label{gsl}
\end{equation}
where the equality holds for the ``equilibrium state'' $\sigma_0$.

Variations $\Delta$ above are finite even though the quantities being varied themselves may diverge. This seems reasonable since one does not expect the ultraviolet behaviour (which is the one responsible for such divergences) to play an important role in these effects of semiclassical origin.

The use of the monotonicity of the relative entropy as in (\ref{ineq}) has been used before by Sorkin \cite{wall2} and Wall \cite{wall} to argue in favour of the GSL, albeit not as rigorously applied to non-stationary spacetimes. Our semiclassical Noether charge approach has the following advantages. First, it is possible to generalise it to other geometrical diffeomorphism-invariant theories of gravity. Second and most importantly, it avoids the assumption that \emph{all} the change in the horizon's area of a dynamical black hole as one moves to the future is caused by the expansion of the existing geodesic congruences generating the horizon, when it is not clear whether or not the emergence of new horizon generators dominates this change in area. This makes this proof distinct of any existing attempts.

\section{Discussion}
This article shows that explicit forms of the zeroth, first, and second laws of black hole thermodynamics emerge from a semiclassical theory of gravity. The meaning of each of their ingredients --- at least in the version of the laws here presented --- was made clearer than ever before since we kept a controlled set of assumptions. In particular, the matter entropy that enters these laws is, per construction, the von Neumann entropy associated with the state restricted to the outside of black holes, as assumed in the literature since the 1970's \cite{Hawking76}. A possible criticism to this framework is the assumption of the existence of consistent semiclassical theories $(M,g,\psi,\rho)$ and the consequent enormous difficulties in applying our results to particular cases since even approximate back reaction effects are already challenging to calculate.

The construction of the semiclassical Hamiltonian as in section II can be applied in broader contexts. For example, Jacobson and Visser \cite{Ted19} recently postulated a semiclassical correction in the first law in causal diamonds\footnote{Strictly speaking, a generalisation of the classical first law using \emph{conformal} Killing fields, which was obtained using the Noether charge methods reviewed in appendix B.}. I believe it is possible to adapt the constructions here to justify their assumption. They also write a relation between the von Neumann entropy as in eq. (\ref{main}), whose right-hand-side represents a source of gravity. The meaning of their variations in expectation values can be assigned with a more precise significance using the definition of the semiclassical Hamiltonian used in this section. Finally, it would be interesting to investigate if there is an analog of the second law (a stronger result than their stationarity condition for the generalised entropy) in causal diamonds. This is an excellent direction to look further into using the methods here developed.

The form of expressing the first and second laws of thermodynamics in the presence of a semiclassical black hole allows the existence of processes that convert ordinary entropy from the exterior of the black hole to its Noether Charge. A prominent example of such a process is the evaporation of black holes, as it emits thermal radiation and loses mass, its Noether-charge entropy decreases while the von Neumann entropy of the radiating fields increases. A crucial hypothesis in deriving eq. (\ref{main}) was that $\rho_0$ is thermal with respect to the Killing parameter (Hartle-Hawking state), but a similar situation already happens in the ordinary first law of thermodynamics, when one expresses heat exchanged to or off the system in terms of its temperature, which is only defined in very special cases. In this sense, equation (\ref{main2}) can be thought of as the first law of thermodynamics written in terms of the heat exchanged rather than in terms of variations in entropy, and remains valid for all physically meaningful states.

The assumption of ``piecewise stationarity'' plays the role of the quasi-steady assumption in the ordinary second law of thermodynamics. In ordinary thermodynamics transitions between different thermodynamic states are thought to be approximated by a succession of equilibrium states. The thermodynamic notion of entropy written in terms of the amount of heat exchanged is only defined under these conditions. Similarly, the notion of the black hole entropy as Noether charge only works under certain approximations, when one can think of the geometry evolving from one stationary state to another, remaining in each stationary stage long enough so that the event horizon can be approximated by a Killing horizon, as discussed above. This approximation is believed to get better the slower the evolution of the geometry.


The derivation of the laws of black hole thermodynamics is consistent with the interpretation of the entropy of a black hole (apart from an additive constant) as accountancy of the amount of information inaccessible its exterior, a view with a strong intuitive appeal that has been discussed since the 1980s \cite{Bombelli}. This interpretation also explains the origins of the ordinary laws of thermodynamics \cite{Jaynes65} without appeal to Boltzmann's microcanonical ensemble, providing unified grounds to both ordinary and black hole thermodynamics. Consequently, while the existence of these laws is the only reason to assign a black hole with an entropy, no further microscopic description of it is needed.

All discussion here was based on changes in entropy. It is possible to speculate that the specification of the absolute value of entropy in terms of the Noether charge will only be possible after something further about quantum gravity is learnt. That would mirror the ordinary thermodynamics since Boltzmann entropy was defined only up to an additive constant before the introduction of the Sommerfeld quantisation rule (and consequently the appearance of the quantum mechanical constant $\hbar$ in his formula). Until then, we can apply and interpret the macroscopic laws of black hole thermodynamics in the semiclassical level as I developed here.

\section*{Acknowledgments}
The author would like to thank his Ph.D. supervisor, Bill Unruh, for hours of discussions that led to the present version of this work, which was partly funded by Conselho Nacional de Desenvolvimento Cient\'ifico e Tecnol\'ogico (CNPq) under the process 200339/2014-1.

\appendix
\section{Geometry of Killing horizons and the zeroth law}
In this appendix we review the basic notions of Killing horizons and its properties and states therein.

Let $(M,g)$ be a spacetime with a globally defined Killing field $\chi^a$ that vanishes in a 2-surface $B$, and whose derivatives across $B$ are nonzero. Then there are a couple of 3-surfaces $\mathfrak{h_+}$ and $\mathfrak{h_-}$ with $\mathfrak{h_+}\cap\mathfrak{h_-}=B$ where $\chi^a$ is null. In this situation $B$ is referred as the \emph{bifurcation surface} and $\mathfrak{h_\pm}$ the \emph{bifurcate horizon} \cite{WaldGR}. Since $g(\chi,\chi)$ is constant along the bifurcate Killing horizon, we must have
\begin{equation}
\nabla^a(\chi^b\chi_b)=-2\kappa\chi^a
 \label{surgra}
\end{equation}
for a function $\kappa$, referred as \emph{surface gravity}. Following \cite{WaldGR, WaldB}, we taking the Lie derivative along $\chi^a$ of both sides of (\ref{surgra}), using the identity $\nabla_a\nabla_b\chi_c=-R_{bcad}\chi^d$ valid for any Killing field, and the skew-symmetry properties of the curvature tensor, it follows that $\mathsterling_\chi\kappa=0$, meaning that the value of $\kappa$ is constant along the orbits of $\chi^a$.

It also follows that
\begin{equation}
    \chi^b\nabla_b\chi^a=\kappa\chi^a,
    \label{intermediate1}
\end{equation}
which can be contracted at each point with a null vector $\ell^a$ defined such that $\chi^a\ell_a=-1$ to obtain
\begin{equation}
 \kappa=-\chi^a\ell^b\nabla_b\chi_a;
 \label{kwl}
\end{equation}

Eq. (\ref{intermediate1}) can multiplied by $\ell^c$ and antisymmetrized to obtain $\ell^{[b}\chi^{c]}\nabla_c\chi^a=\kappa\epsilon^{ab}$, where $\boldsymbol\epsilon$ is the volume form on $B$, and the antisymmetry between $b$ and $c$ on the left hand side is guaranteed by the Killing equation already, so the brackets can be removed. If one completes the $(\ell,\chi)$ basis to obtain a Newman-Penrose basis and expresses the inverse metric in terms of direct products of this vector basis, and using the definition of $\kappa$,
\begin{equation}
 -\nabla^b\chi^a=\kappa\epsilon^{ab}.
 \label{volformB}
\end{equation}

The above are standard results for Killing horizons\cite{WaldGR, WaldB}. The imply a version of the zeroth law of black hole mechanics as follows: 

The square of (\ref{volformB}) gives $-2\kappa^2=(\nabla_a\chi_b)(\nabla^a\chi^b)$. Taking a derivative of this along a direction $t^a$ tangent to $\mathfrak h_\pm$ and orthogonal to $\chi^a$,
\[2\kappa t^a\nabla_a\kappa=-t^c\left(\nabla_c\nabla_a\chi_b\right)(\nabla^a\chi^b)=t^cR_{abc}^{\ \ \ d}\chi_d\nabla^a\chi^b=0,\]
where the Killing identity and the symmetry properties of the curvature tensor were used.

For a \emph{bifurcate} horizon, this means $t^a\nabla_a\kappa=0$. This means that $\kappa$ is not only constant both along the generators of the horizon, but also across different generators, this result is a version of \emph{the zeroth law of black hole mechanics}. This version, differently from the version in \cite{bch}, does not depend on any assumption regarding the dynamics of the theory as long as it is described by a Killing horizon with $\kappa\neq0$.

We now provide a simple adaptation of arguments originally presented in \cite{unruh} to exhibit a sufficient condition for the existence of a thermal state, as we assumed in the main text. A stronger and more general result using different methods can be found in \cite{KayWald}. 

We introduce the coordinates $(u,\varrho,x^\alpha)$ where $\chi=\partial/\partial u$, $\ell=\partial/\partial\varrho$ with $\varrho=0$ at $\mathfrak{h_\pm}$ and $x^\alpha$ are the remaining coordinates.

In these coordinates, with the abbreviations $F=-\chi^a\chi_a$ and $g_{u\alpha}=\chi_a(\partial/\partial x^\alpha)^a$ the metric can be written as
\[\ud s^2=-F\ud u^2-2\ud u\ud\varrho+2g_{u\alpha}\ud u\ud x^\alpha+g_{\alpha\beta}\ud x^\alpha\ud x^\beta,\]
where none of the metric coefficients depend on $u$, since it is the Killing parameter. Thus, only keeping terms up to the first order in $\varrho$ and using
\[\frac{\ud F}{\ud\varrho}=\ell^b\nabla_b(-\chi^a\chi_a)=-2n^b\chi^a\nabla_b\chi_a=2\kappa,\]
near the horizon where $F=0$,
\[\ud s^2=-2\kappa\varrho\ud u^2-2\ud u\ud\varrho+2g_{u\alpha}\ud u\ud x^\alpha+g_{\alpha\beta}\ud x^\alpha\ud x^\beta.\]
The change of coordinates from $(u,\varrho)$ to $(t,r)$ with $t\equiv u+\int F^{-1}\ud\varrho$ and $r=\sqrt{2\varrho/\kappa}$ brings the line element to the form
\begin{equation}
 \ud s^2=-\kappa^2 r^2\ud t^2+\ud r^2+2g_{u\alpha}\ud u\ud x^\alpha+g_{\alpha\beta}\ud x^\alpha\ud x^\beta,
\end{equation}
whose $(t,r)$ section is identical to the Minkowski spacetime's in Rindler coordinates.

If besides $\chi^a$ we suppose $(M,g)$ admits a globally defined timelike Killing field $\xi^a$, the restriction of the vacuum state associated with $\xi^a$ to a spacetime wedge like $I^+(\mathfrak{h_-})\cap I^-(\mathfrak{h_+})$ will be a thermal state with respect to the Hamiltonian conjugate to the orbits of $\chi^a$, since to evaluate the density matrix obtained by such a restriction it is enough to know the value of the fields at the bifurcated horizon, following the same steps as in \cite{unruh}.

\section{Hamiltonian formalism and the Noether charge in classical theory}
This appendix summarizes the findings from \cite{Iyer} that are relevant in the main text.

Starting from the variational principle for a classical diffeomorphism-covariant theory over an $n$-dimensional manifold $M$, the action $I_0=\int_M\mathbf L$ is the integral of the Lagrangian $n$-form $\mathbf L$, which is a function of the matter fields and spacetime metric, collectively denoted by $\phi$, and their derivatives. The variation of the Lagrangian can be generally expressed as
\begin{equation}
\delta\mathbf L=\mathbf E\cdot\delta\phi+\ud\boldsymbol\Theta(\phi,\delta\phi),
\label{deltaL}
\end{equation}
 where $\boldsymbol\Theta$ is a $(n-1)$-form linear in $\delta\phi$ and its derivatives, referred to as \emph{sympectic potential}. The second term is produced after ``integration by parts'' and can be converted, via Stoke's theorem, into a boundary term. Thus, the equations of motion are recognised as $\mathbf E=0$.
 
 The potential $\boldsymbol\Theta$ can naturally be seen as a one-form in configuration space $\mathscr F$, where field variations $\delta\phi$ are viewed as vectors in its tangent space $T_\phi\mathscr F$. Then we can produce a two-form $\boldsymbol\omega$ in this space by taking its exterior derivative. I will reserve the symbol $\ud$ for exterior derivatives in spacetime, and simply write instead
\begin{equation}
\boldsymbol\omega(\phi,\delta_1\phi,\delta_2\phi)=\delta_2\boldsymbol\Theta(\phi,\delta_1\phi)-\delta_1\boldsymbol\Theta(\phi,\delta_2\phi),
\label{sfgen1}
\end{equation}
which is a closed and antisymmetric $(n-1)$-form in spacetime. The particular case
\begin{equation}
\boldsymbol\omega_\chi\equiv\boldsymbol\omega(\phi,\delta\phi,\mathsterling_\chi\phi).
\label{sfgen2}
\end{equation}
will be extremely useful. This $(n-1)$-form in spacetime is ready to be integrated over a $(n-1)$-surface $C$, which we choose to be a partial Cauchy surface.
\begin{equation}
 \Omega(\phi,\delta_1\phi,\delta_2\phi)=\int_C\boldsymbol\omega(\phi,\delta_1\phi,\delta_2\phi),
 \label{symp}
\end{equation}

We wish to make this 2-form in $\mathscr F$ a symplectic form. This requires it to be closed and non-degenerate. By construction, $\Omega$ is exact and consequently closed, but it is degenerate. For example, $\Omega_{AB}\left(\delta\phi\right)^B$ is zero for all non-zero $\left(\delta\phi\right)^A$ whose restriction over $C$ happens to vanish\footnote{Here I use the abstract index notation with capital letters to represent tensor arguments of $\mathscr X(\mathcal P)$, i.e. vector fields in phase space, and with lower case symbols for arguments in $\mathscr X(M)$.}.

However, if one follows the prescription introduced by Lee and Wald \cite{LeeW} called the symplectic quotient, it is possible to define a proper phase space $\mathcal P$ of the theory by setting an appropriate quotient of $\mathscr F$ so that $\mathcal P$ is automatically equipped with a symplectic 2-form $\Omega_{AB}$, induced by the 2-form in phase space $\boldsymbol\omega_\chi$, which admits an inverse. The prescription goes roughly as follows.

Let $\eta^A$ and $\zeta^A$ denote degenerate fields, i.e., $\Omega_{AB}\eta^B=\Omega_{AB}\zeta^B=0$. Then the (phase space) commutator $\psi^A=[\eta,\zeta]^A$ is also degenerate. Indeed,
\[\Omega_{AB}\psi^B=\Omega_{AB}\mathsterling_\eta\zeta^B=\mathsterling_\eta(\Omega_{AB}\zeta^B)-\zeta^B\mathsterling_\eta\Omega_{AB}=0,\]
where the first term is zero by hypothesis, and the second is zero as a consequence of Cartan's formula $\mathsterling_\eta\Omega_{AB}=\eta^C(\ud\Omega)_{CAB}+(\ud(\boldsymbol\Omega\cdot\boldsymbol\eta))_{AB}=0$. Consequently, by Frobenius' theorem \cite{lee} the subbundle of $T\mathscr F$ consisting of the degeneracy subspaces of $\Omega_{AB}$ admits an integral submanifold. Then one defines an equivalence relation $\psi\sim\phi$ of $\mathscr F$ if $\psi$ and $\phi$ lie on the same integral submanifold. As a set, the phase space $\mathcal P$ is the quotient of $\mathscr F$ by this relation. To endow it with a symplectic structure, one uses the projection $\varpi:\mathscr F\to\mathcal P$ to identify points on the manifolds (i.e. $\varpi$ maps every element of $\mathscr F$ to its equivalence class in $\mathcal P$) and its pullback to identify vectors on the tangent space. The symplectic form on $\mathcal P$ defined so that
\[\Omega_{AB}\ \left(\varpi^*\psi^A\right)\left(\varpi^*\phi^B\right)=\Omega_{AB}\psi^A\phi^B, \quad\forall\ \psi,\phi\in\mathscr F,\]
where $\Omega_{AB}$ on the left-hand side denotes, with a small abuse of notation the symplectic form on $\mathcal P$.

In possession of a symplectic manifold $(\mathcal P,\Omega_{AB})$, the missing ingredient for making classical mechanics is a Hamiltonian function $H:\mathcal P\to\mathbb R$ generating a desired flow. If $\varphi$ and $\varphi+\delta\varphi$ satisfy the equations of motion $\mathbf E=0$, then $\ud\boldsymbol\omega_\chi=0$ and $\Omega_\chi$ is independent of the choice of the Cauchy surface $C$, so $\mathsterling_\tau\Omega_\chi=0$, where $\tau^A=(\mathsterling_\chi\varphi)^A$ represents the time evolution in the solution subspace of $\mathcal P$. As any subspace, it is simply connected, implying there exists a $H_\chi$ such that
\begin{equation}
(\ud_{\mathcal P} H_\chi)_A=(\Omega_\chi)_{AB}\ \tau^B, 
\label{dhp}
\end{equation}
where $\ud_{\mathcal P}$ represents the exterior derivative in that space (which coincides with the differential since $H_\chi$ is a function). Or, in our customary notation, a Hamiltonian conjugate to the parameter along the orbits of $\chi^a$ must satisfy
\begin{equation}
\delta H_\chi=\Omega(\phi,\delta\phi,\mathsterling_\chi\phi)\equiv\Omega_\chi.
\label{dh}
\end{equation}

Eq. (\ref{dh}) is equivalent to
 $$\left(\mathsterling_\chi\varphi\right)^A=(\Omega^{-1})^{AB}(\nabla H_\chi)_B,$$ the usual canonical equations in the dynamically accessible phase space \cite{Arnold}. Indeed, when we can apply the inverse symplectic form to eq. (\ref{dhp}), we obtain the familiar canonical equations $\tau^A=\Omega_\chi^{AB}(\ud_{\mathcal P} H)_B$.

The Lagrangian is covariant under diffeomorphisms, its Lie derivative is determined by the Lie derivative of the fields $\phi$, $$\mathsterling_\chi\mathbf L=\frac{\partial \mathbf L}{\partial\phi}\mathsterling_\chi\phi,$$ then the $(n-1)$-form
\begin{equation}
    \mathbf j_\chi=\boldsymbol\Theta(\phi,\mathsterling_\chi\phi)-\chi\cdot\mathbf L
\label{defj}
\end{equation}
is closed whenever the equations of motion (both on the matter fields and on the metric) are satisfied, more explicitly,
\begin{multline}
\ud\mathbf j_\chi=\mathsterling_\chi\mathbf L-\mathbf E\cdot\mathsterling_\chi\phi-\ud(\chi\cdot\mathbf L)=\\
=\mathsterling_\chi\mathbf L-\mathbf E\mathsterling_\chi\phi-\chi\cdot\ud\mathbf L-\mathsterling_\chi\mathbf L=\\
=-\mathbf E\cdot\mathsterling_\chi\phi=0,
\label{djay}
\end{multline}
where the identity $\chi\cdot\mathbf L=\mathsterling_\chi\mathbf L-\ud(\chi\cdot\mathbf L)$, valid for any form $\mathbf L$\cite{lee} was used in the second step, and that $\ud\mathbf L=0$, since $\mathbf L$ is a top form.

For a simply connected domain of $\mathbf j_\chi$, as it will be the case for a spacetime wedge bounded by a bifurcated Killing horizon like $I^+(\mathfrak{h_-})\cap I^-(\mathfrak{h_+})$ this means that $\mathbf j_\chi$ is also exact, $\mathbf j_\chi=\ud\mathbf Q_\chi$. Taking a variation of $\mathbf j_\chi$ and using the covariant property of $\boldsymbol\Theta$, we relate $\mathbf j_\chi$ and $\boldsymbol\omega_\chi$:
\begin{multline}
\delta\mathbf j_\chi=\delta\boldsymbol\Theta(\phi,\mathsterling_\chi\phi)-\chi\cdot\delta\mathbf L=\\
=\delta\boldsymbol\Theta(\phi,\mathsterling_\chi\phi)-\chi\cdot\mathbf E\cdot\delta\phi-\chi\cdot\ud\boldsymbol\Theta(\phi,\delta\phi)=\\
\delta\boldsymbol\Theta(\phi,\mathsterling_\chi\phi)-\mathsterling_\chi\boldsymbol\Theta(\phi,\delta\phi)+\ud(\chi\cdot\boldsymbol\Theta(\phi,\delta\phi))-\chi\cdot\mathbf E\cdot\delta\phi\\
\Rightarrow\boldsymbol\omega_\chi=\delta\mathbf j_\chi-\ud(\chi\cdot\boldsymbol\Theta)+\chi\cdot\mathbf E\delta\phi.
\label{j}
\end{multline}

We can transform $\delta H_\chi$ for solutions into a pure boundary term using eqs. (\ref{j}) and (\ref{dh}) and applying Stokes' theorem.

In order $H_\chi$ to be defined, we assume the necessary and sufficient condition that there exists a $(n-1)$-form $\mathbf B$ such that\footnote{If this form does not exist, we cannot define a Hamiltonian for the field $\chi^a$. $\mathbf B$ is known \cite{Iyer} to exist in general relativity when $\chi^a$ is an assymptotic time translation under reasonable behaviours of the matter fields at infinity.} 
\begin{equation}
\delta\int_{\partial\Sigma}\chi\cdot\mathbf B=\int_{\partial\Sigma}\chi\cdot\boldsymbol\Theta.
\label{intB}
\end{equation}
Using eqs. (\ref{dh}-\ref{j}) whenever the equations of motion are obeyed,

\begin{equation}
 H_\chi=\int_{\partial\Sigma}\mathbf Q[\chi]-\chi\cdot\mathbf B.
 \label{hhat}
\end{equation}

This Hamiltonian (\ref{hhat}) is only defined up to a constant, since the equations of motion only impose requirements over its variation, not on its absolute value.

If such a form $\mathbf B$ exists, then instead of the original action $I_0$, we adopt the modified action
\begin{equation}
I=\int_M \mathbf L-\int_{\partial M}\mathbf B,
 \label{action}
\end{equation}
which generates the same equations of motion as $I_0$, and has the advantage that $I$, and not $I_0$, is extremised when we keep the boundary of spacetime fixed, which is of great importance in approaches to quantum gravity, and crucial for the so-called Euclidean methods to compute gravitational entropy. In general relativity, i.e., when $\mathbf L$ is the Einstein-Hilbert Lagrangian, $\mathbf B$ on the boundary is proportional to the trace of the extrinsic curvature of $\partial M$ times the induced volume form.
\begin{equation}
\mathbf L=\frac{\mathscr R}{16\pi}\boldsymbol\epsilon+\frac{1}{2}g_{ab}\nabla^a\psi\nabla^b\psi\ \boldsymbol\epsilon,
\label{msfmcgr}
\end{equation}
where $\mathscr R$ is the scalar curvature and $\boldsymbol\epsilon$ is the volume element on the spacetime. For this Lagrangian we obtain
\begin{equation}
\begin{gathered}
\langle\Theta^\psi_{abc}\rangle=\epsilon_{abcd}\nabla^d\langle\psi\delta\psi\rangle,\\
\Theta^g_{abc}=\frac{1}{16\pi}\epsilon_{dabc}g^{de}g^{fh}(\nabla_f\delta g_{eh}-\nabla_e\delta g_{fh}),\\
(j^\psi_\chi)_{abc}=\epsilon_{dabc}\langle\nabla_e\psi\chi^{e}\nabla^{d}\psi\rangle-\frac{1}{2}\epsilon_{eabc}\langle\nabla^d\psi\nabla_d\psi\chi^e\rangle,\\
(j^g_\chi)_{abc}=\frac{1}{8\pi}\epsilon_{dabc}\nabla_e\nabla^{[e}\chi^{d]}.
\end{gathered}
\label{msfmcgr2}
\end{equation}

Integrating eq. (\ref{dh}) over a Cauchy surface $C$ and using the second of (\ref{msfmcgr2}), we can write the $g$-part of the varied Hamiltonian in the ADM form \cite{MTW, WaldGR}
\[-\frac{1}{32\pi}\int_C\delta h_{ab}\mathsterling_\chi p^{ab}-\mathsterling_\chi h_{ab}\delta p^{ab},\]
where $h_{ab}$ is the induced metric over $C$, and $p^{ab}=\sqrt{\det h}(K^{ab}-h^{ab} K^c_c)$ is its conjugate momentum with $K_{ab}$ the extrinsic curvature of $C$.

The $\psi$-part is easily identified --- using  the first and third of (\ref{msfmcgr2}) --- with
\[\delta\int_C \langle T_{df}\rangle\chi^f\epsilon^d_{\ abc}, \quad T_{ab}=\nabla_a\psi\nabla_b\psi-\frac{g_{ab}}{2}\nabla^c\psi\nabla_c\psi,\]
which is in the familiar form for the Hamiltonian derived from the field equations.

The same Hamiltonian is generated by the dynamics of $G_{ab}=8\pi\langle T_{ab}\rangle$, the usual semiclassical Einstein's equations.


\end{document}